# Rydberg Exciton Dynamics in the Blockade Regime of Cu$_2$O

Gillian E. Minarik[1,†], Eric A. Arsenault[1,†,*], Vinícius da Silveira Lan Avelar[1], Taketo Handa[1], and X.-Y. Zhu[1,*]

[1] Department of Chemistry, Columbia University, New York, NY 10027, USA

**Abstract:** Hosting giant Rydberg excitons with principal quantum numbers up to $n = 30$, cuprous oxide (Cu$_2$O) provides a rare solid-state setting for exploring Rydberg physics, as exemplified by the blockade effect. Here, we access the strongly interacting regime at high excitation densities ($10^{14}$-$10^{16}$ cm$^{-3}$) and resolve the corresponding blockade dynamics for $n = 2$-7 using time-resolved spectroscopy. We find that Rydberg blockades are primarily governed by resonant dipolar interactions and that exciton recombination is coupled to the blockade effect itself. These findings demonstrate the potential for manipulating Rydberg excitons in the strongly interacting blockade regime in a solid-state system.

[†] These authors contributed equally.
[*] To whom correspondence should be addressed. EAA: eaa2181@columbia.edu; XYZ: xyzhu@columbia.edu.

The ability to generate and control strongly interacting quantum systems is essential for applications across quantum information processing, sensing, and quantum optics. With giant wavefunction extensions promoting huge dipolar interaction effects, excited Rydberg atoms and ions are leading platforms in this domain [1–3]. These long-range dipolar interactions between Rydberg atoms lead to an efficient blockade effect—a phenomenon in which one excited state "blocks" the excitation of another in its vicinity by detuning its excitation energy [3–5]. This effect has become increasingly central to enabling precise qubit interactions and control in emerging quantum computing platforms [6–9]. Generally, Rydberg systems include any hydrogen-like system featuring highly excited states with energies following the Rydberg formula, $E_n = -\frac{R_H}{(n-\delta)^2}$, where $n$ is the principal quantum number, $R_H$ is the Rydberg constant, and $\delta$ is the quantum defect [10,11].

Rydberg atoms find their solid-state analogs in excitons, the fundamental optical excitations of semiconductors. In principle, Wannier excitons follow the Rydberg formula, but higher excitonic Rydberg states are seldom observed. Rydberg excitons up to $n = 5$ have been measured in transition metal dichalcogenides [12] but their spectral features are very weak; this likely owes to a combination of factors such as high recombination rates [13,14], and a broad, optically bright 1s exciton background. The discovery of giant Rydberg excitons in $Cu_2O$ provides a rare and valuable example of a semiconductor system that exhibits nearly ideal Rydberg physics [15,16]. Experiments on $Cu_2O$ have revealed Rydberg excitons with energies described by the Rydberg formula at a small $\delta$ (= 0.08) [17], reaching principal quantum numbers as high as $n = 30$ [16]. These observations may be attributed to favorable material properties, including but not limited to, extremely narrow linewidths [18] and the absence of the 1s exciton background [17]. Remarkably, these excitons also obey a series of characteristic scaling laws with respect to *n [19]*, a defining feature of Rydberg systems [20,21]. $Cu_2O$ thus not only supports the formation of high $n$ excitons but also enables exploration of fundamental many-body driven phenomena, such as the Rydberg blockade, in a solid. Of relevance to the blockade effect are the scaling laws for the Rydberg exciton radius and corresponding dipole moment ($\propto n^2$), as well as the polarizability ($\propto n^7$), which govern the resonant Förster-type and non-resonant van der Waals (vdW) interactions, respectively [17,20–23]. Experimental evidence for both resonant and non-resonant blockade mechanisms has been reported for Rydberg excitons in $Cu_2O$ [15,17,22,23], although discerning their relative contributions as a function of $n$ remains a challenge. Moreover, the temporal evolution of blockade effects and their influence on exciton population dynamics are not well-understood.

The transient nature of excitons and their interactions fundamentally distinguish the Rydberg states in $Cu_2O$ from those of atomic systems. Excitonic Rydberg states rely on injected excitons that themselves exist only transiently before recombination. Despite this dynamic nature, relatively few time-resolved studies have been conducted on Rydberg excitons in $Cu_2O$ [23–27]. To date, most measurements on $Cu_2O$ have been carried out under continuous-wave (CW) and quasi-equilibrium conditions [22]. Direct time-domain measurements of Rydberg exciton coherence times in $Cu_2O$ only recently emerged for the *s*-series, [24]. Panda et al. were able to temporally distinguish μs decay mechanisms using a two-color pump-probe approach, assigning distinct timescales to plasma relaxation, impurity dynamics, and para-exciton lifetimes [27]. These works demonstrate the promise of time domain approaches to disentangle the complex decay dynamics in $Cu_2O$, which notably deviate from the well-known radiative lifetime scaling $\propto n^3$ for Rydberg states [15,16,19,21]. Expected to occur on ultrafast timescales [27], many-body excitonic interactions such as the Rydberg blockade remain unexplored in the time-domain. To truly realize $Cu_2O$ as a platform for manipulating Rydberg states, it is imperative to understand how these dipolar interactions affect exciton dephasing and decay processes, particularly at high densities, and how blockade dynamics unfold more generally.

Here, we apply femtosecond pump-probe spectroscopy to investigate blockade dynamics in $Cu_2O$. In this approach, the pulsed excitation drives the system into a high-density regime beyond what is typically achieved in a CW experiment. We resolve Rydberg states up to $n = 7$ and achieve exciton densities in the range $10^{14}$-$10^{16}$ cm$^{-3}$, thereby accessing Rydberg blockade dynamics over a broad interaction regime. Our time-resolved measurements provide evidence that resonant dipole interactions likely govern the blockade dynamics for states up to $n = 7$. By systematically varying the excitation fluence, we directly tune the strength of Rydberg blockade interactions and uncover a strong correlation between the photobleaching (i.e., exciton population) and photoinduced absorption (i.e., blockade) dynamics for each observed $n$, suggesting that exciton recombination is governed by the blockade. More generally, these dynamics also reveal that the observed Rydberg excitons have unexpectedly long lifetimes on the order of 10s to 100s of ps, which are also sustained even in a high-excitation, highly interacting regime. These findings illustrate the critical role of many-body blockade effects in shaping exciton dynamics in $Cu_2O$.

Spectral Signatures of Rydberg Blockades. For the ultrafast pump-probe measurements, the differential transmission signal, $\Delta T/T$, was collected as a function of pump-probe delay, $\Delta t$. Here, $T$ denotes probe transmission without the pump and $\Delta T$ denotes the pump-induced change in transmission (see details in Appendix A). The above-gap pump excitation ($h\nu_1 = 2.33$ eV) rapidly decays on a sub-ps timescale [22,28] to populate the Rydberg exciton manifold of the yellow $p$-series of $Cu_2O$ [15,17]. A probe pulse ($h\nu_2 = 2.11$-$2.18$ eV) spanning the $p$-series spectral region monitors the corresponding Rydberg exciton dynamics with sufficient energy resolution to resolve states up to $n = 7$, while retaining sub-ps time resolution. The resulting transient response was measured as a function of pump fluence, which was varied over two orders magnitude ($f = 0.07$-$8.15$ $\mu J/cm^2$), to ensure that the blockade regime was achieved even at the lowest observable states. Fig. 1a-c show representative pseudocolor plots of the spectrally and temporally resolved data at 7 K at three selected pump fluences of $f = 0.07$, $1.01$, and $8.15$ $\mu J/cm^2$, corresponding to estimated overall exciton densities of $n_{ex} = 4.1 \times 10^{14}$, $5.9 \times 10^{15}$, and $4.9 \times 10^{16}$ $cm^{-3}$, respectively. See Fig. B1 for the data at other excitation densities. In the pump-probe maps, negative $\Delta T/T$ (blue) represents photoinduced absorption (PIA), while positive $\Delta T/T$ (red) represents photobleaching (PB).

The structure of the transient $\Delta T/T$ spectrum crucially encodes information about inter-Rydberg exciton interactions. For better visualization of these effects (as well as the higher-lying states at $n = 6$ and 7), we extract fluence-dependent line cuts at early delay ($\Delta t = 1$ ps, Fig. 1d) and time-dependent cuts at a fixed pump fluence ($f = 3.06$ $\mu J/cm^2$, Fig. 1e). To interpret the spectral features, we first consider the system in the absence of pump excitation. In this case, the probe transmission spectrum exhibits absorption features at energies determined by the Rydberg formula (red line, Fig. 1f). When the pump is on, repulsive inter-exciton interactions, $V(r)$, introduce an energetic cost to the optical excitation of probe excitons near those created by the pump. This leads to a blue shift of the probe resonance, giving rise to a Rydberg blockade [17,22] (blue line, Fig. 1f). Spatially, this interaction range is defined by the blockade radius, $r_B$, (on the order of 100s nm to few $\mu m$) [17,22], corresponding to the distance at which the Rydberg state is detuned by the linewidth of its unperturbed resonance. The resulting differential transmission generates a distinct derivative-like $\Delta T/T$ pair (purple line, Fig. 1f) for each Rydberg level $n$ that experience blockade interactions: a positive $\Delta T/T$ near the resonance corresponds to a PB, while a blue-shifted negative $\Delta T/T$ indicates a blockade-induced PIA. This derivative-like structure is observed at each Rydberg exciton level from $n = 2$-$7$, confirming that the blockade regime is reached across the entire accessible $n$ range. Finally, the strength of the blockade effect for each $n$, reflected in the amplitude of the corresponding PIA feature, increases with excitation density (i.e., fluence), as expected.

Evident also in Fig. 1d for each $n$ is the presence of an isosbestic point—where the PB and blue-shifted PIA cross at the same probe energy, independent of excitation fluence. Using a CW pump-probe approach, Heckötter et al. previously established that such an isosbestic point is a characteristic spectral manifestation of Rydberg blockade effects which can be well-described by a two-body spatial correlation function [22]. In the current measurements, the persistence of this spectral structure throughout the measured time window (Fig. 1e) confirms the sustained presence of the photoinduced Rydberg blockade across all observed $n$, which is already present just after the excitation. Note that such derivative-like spectra are often attributed to phase-space filling in pump-probe experiments on conventional semiconductors, which would typically result in a red-shifted PIA feature due to screening and band renormalization [29]. In $Cu_2O$, by contrast, the PIA appears on the blue side of the exciton resonance, clearly indicating the distinct repulsive nature of the blockade interaction.

Rydberg Exciton Blockade Dynamics. We now turn to the temporal evolution of the PB and PIA features. PBs reflect the population dynamics of the Rydberg excitons, whereas PIAs capture the dynamics directly associated with the Rydberg blockade. The fluence-dependent time traces of the PB (red) and PIA (blue) for each Rydberg level are compared in Fig. 2. These time traces are taken at the probe energies where $|\Delta T/T|$ is maximized for each feature, with light to dark shaded lines indicating increasing fluence (in the range $f = 0.07$-$8.15$ µJ/cm$^2$). Fig. 2 reveals two striking observations regarding the extended timescales observed across the range $n = 2$-$7$: i) for $\Delta t > 0$, the Rydberg exciton PB and PIA features persist for 10s to 100s of ps; ii) for $\Delta t < 0$, the "probe-pump" signal persists for 100s of fs to few-ps. While the positive-delay ($\Delta t > 0$) signal reflects exciton population dynamics (PB) and dynamic blockade interactions (PIA), the negative-delay ($\Delta t < 0$) response mainly reveals dephasing dynamics. We focus on the $\Delta t > 0$ behavior in the following and discuss $\Delta t < 0$ towards the end.

At $\Delta t > 0$, the Rydberg excitons not only display generally long-lived signals but also exhibit a clear $n$-dependent trend: both PB and PIA dynamics extend to longer timescales as $n$ increases.

Furthermore, for each observed Rydberg state, the PB and PIA exhibit nearly the same temporal structure and fluence dependence, indicating a shared underlying origin. More specifically, the symmetry between the PB transient for a given $n$ and the nearest blue-shifted PIA—associated with detuned excitons of the same $n$ (Fig. 1f)—strongly suggests that a resonant Förster-type blockade mechanism dominates for $n \leq 7$ in these experiments [17]. If, on the other hand, the blockade arose from interactions between excitons at $n$ and a different Rydberg level $n'$, the PB recombination dynamics would instead mirror those of a more spectrally shifted PIA feature.

To disentangle these dynamics and extract associated time constants, we employ Lifetime Density Analysis (LDA) using an open-source software [30]. Unlike approaches more widely employed to extract kinetic information, LDA is model-agnostic and excels at resolving overlapping processes on multiple timescales in complex systems with many closely spaced states [31]. As detailed in Appendix A, LDA fits transient data using hundreds of weighted exponentials and returns a continuous distribution of time constants (i.e., lifetimes). The amplitude of the distribution reflects the relative contribution of each time constant to the dynamics. Because LDA returns a continuous distribution of time constants, peaks in the LDA amplitude spectrum correspond to dominant kinetic pathways—either signal growth or recombination—and can therefore also naturally handle the non-monotonic $\Delta t > 0$ dynamics observed for $n > 3$.

Fig. 3a shows a typical LDA map for a pump fluence of 8.15 µJ/cm$^2$; results at other pump-fluences are presented in Fig. B2. To focus on the blockade dynamics, Fig. 3b-f presents lifetime distribution line cuts for the PB (purple) and PIA (green) features of each Rydberg level. The fluence-dependent LDA results are shown via shaded lines, where increasing fluence is indicated by darker color. We note that $n = 7$ is omitted from further analysis due to lower signal-to-noise. Importantly, PB and PIA features in the LDA results follow distinct sign conventions that indicate growth or recovery (i.e., decay) of the excited state population. For the PB (purple, Fig. 3), a positive amplitude indicates recovery, while a negative amplitude indicates growth. Conversely, for the PIA (green, Fig. 3), a positive amplitude corresponds to growth, and a negative amplitude corresponds to recovery. Again, peaks in these distributions identify dominant dynamical pathways.

For the Rydberg exciton states analyzed in this way ($n = 2$-6), a clear symmetry between the PB and PIA distributions is observed, consistent with the trends in Fig. 2. This provides quantitative confirmation that a resonant Förster-type blockade mechanism [17] governs the dynamics in this regime: the exciton recombination time for a given *n* shares identical time constants only with its corresponding detuned blockade feature. This relationship is further supported by the lifetime density map in Fig. 3a, where the full spectral extent of each blockade region exhibits characteristic time constants that align exclusively with the PB dynamics of the same *n*. These time constants also exhibit similar fluence (i.e., exciton density) dependence which directly tunes the strength of Rydberg blockade interactions. Because the PB time constants track the fluence-dependent PIA time constants so closely, the data suggests that exciton population dynamics (monitored via PB) is fundamentally dictated by the blockade interactions (monitored via PIA) for each *n*. In addition to these main lifetime peaks, PB growth peaks with faster time constants (red arrows, Fig. 3) and accompanying PIA peaks, are also resolved and likely reflect population transfer from higher *n* states in the manifold.

Given that the PB and PIA features exhibit essentially identical characteristic timescales for a given *n*, we focus on the exciton recombination time constants, $\tau_r$, in the following. Fig. 4 presents $\tau_r$ as a function of *n* (see Fig. B3 for the blockade recovery time constants, $\tau_b$, as a function of *n*). As shown by the dynamics discussed above, exciton lifetime slows significantly with increasing *n*, consistent with the *n*-dependent scaling of blockade interaction strength. This behavior, however, stands in sharp contrast to the exciton recombination dynamics reported in other semiconductors, including alternative Rydberg exciton hosts such as TMDs, where decay times for Rydberg excitons are typically much shorter than those of the lowest-lying excitonic state [13,14]

The exciton recombination time exhibits a decrease with increasing exciton density (Fig. 4), as also captured by the fluence-dependent peak shifts in Fig. 3. While LDA does not assume a specific kinetic mechanism, it does model the signal as a continuous distribution of exponential decays, which may not fully be able to capture second (or higher) order contributions to blockade dynamics. Consequently, the extracted time constants retain a residual exciton density dependence (i.e., at every *n*, the time constants decrease with increasing exciton densities). Yet more intriguingly, the recombination time, $\tau_r$, saturates once the pump fluence exceeds ~4 μJ/cm$^2$, with all high-fluence data collapsing onto a single *n*-dependent curve in Fig. 4. Such saturation with respect to exciton density points to a dominant two-body (or more generally few-body) blockade mechanism, in agreement with the isosbestic structure described above and in Ref. [22]. This saturation behavior is in stark contrast to other many-body dynamical behaviors in more conventional setting, e.g., seen in the bimolecular recombination regime for highly excited semiconductors where the exciton recombination becomes continuously faster with increasing excitation density with no saturation behavior [32]. Further, in addition to this fundamentally intriguing behavior, the sustained lifetimes in the strongly interacting regime may also find relevance for their dynamical control of Rydberg excitons. A detailed theoretical treatment that explicitly includes multi-exciton correlations will facilitate full exploration of this trend and understanding.

In addition to the longer-time dynamics for $\Delta t > 0$, substantial $\Delta t < 0$ signal is observed (Figs. 1a-c). The latter arises from coherent interference between the pump-induced polarization and the free induction decay of the Rydberg excitons resonantly excited by the preceding probe pulse [33]. While it is difficult to extract precise time constants, the pre-time-zero tail clearly becomes longer with increasing $n$ (from ~sub-ps to ~few-ps), indicating an increase in the dephasing time for increasing $n$, consistent with the Rydberg scaling for the linewidth [15–17]. These relatively long timescales provide further insight into the practical utility of Rydberg excitons for quantum control or coherent manipulation. As a final note, the more significant spectral interference observed at higher $n$ results from the more closely spaced exciton resonances.

Concluding Comments. We performed time-resolved measurements on $Cu_2O$ with sub-ps temporal resolution and energy resolution capable of resolving Rydberg exciton dynamics up to $n = 7$. These experiments reveal the $n$-dependent transient behavior of Rydberg excitons in the blockade regime, offering insight into carrier dynamics essential for developing strongly interacting solid-state systems inspired by atomic and ionic Rydberg arrays. For $n \leq 7$, the data point to resonant (Förster-type) interactions as the dominant blockade mechanism. Fluence-dependent traces show nearly identical kinetics for exciton recombination (PB) and blockade-induced absorption (PIA), suggesting that the blockade governs the recombination dynamics. Further, the recombination time lengthens systematically with increasing principal quantum number, $n$, consistent with dipolar interactions that polarize each exciton in the presence of its neighbors. From a mutual screening perspective, the electron and the hole in a polarized exciton become no longer spatially co-localized, slowing the recombination. Given the scaling laws of the Rydberg exciton dipole moment ($\propto n^2$) and polarizability ($\propto n^7$), we expect the trend observed here for $n = 2$-$6$ (Fig. 4) to continue even more dramatically for higher $n$. The combination of long excitonic lifetime and blockade interaction times, along with consistently extended dephasing times, all of which scale with *n*, makes $Cu_2O$-hosted Rydberg excitons a compelling platform for realizing strongly interacting solid-state Rydberg systems.

## Acknowledgements

The spectroscopic work was supported by Department of Defense Multidisciplinary University Research Initiative (MURI) grant number W911NF2410292. Methodology development and theoretical analysis were supported by Department of Energy Office of Basic Energy Sciences (DOE-BES) under award DE-SC0024343. The authors acknowledge the use of facilities and instrumentation supported by the Columbia Nano Initiative, and the NSF Materials Research Science and Engineering Center DMR-2011738. EAA gratefully acknowledges support from the Simons Foundation as a Junior Fellow in the Simons Society of Fellows (965526). We thank Professor Nobuko Naka of Kyoto University for invaluable advice on sample preparation, Dr. Amir Zangiabadi for assistance in polishing the crystals, and Dr. Daniel G. Chica and Professor Xavier Roy for assistance in x-ray diffraction to determine the crystal orientation.

## Competing Interests

The authors declare no competing interests.

## Data Availability Statement

The data that supports the findings of this article are openly available [34].

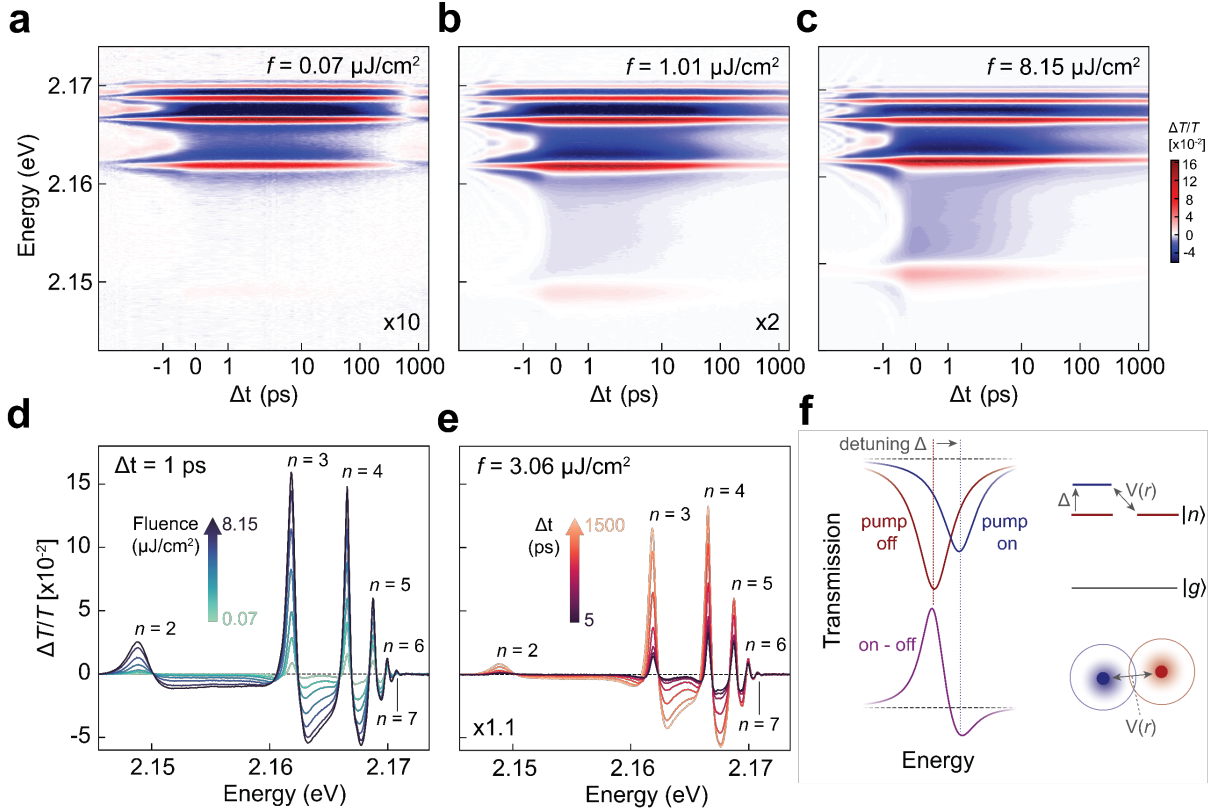

**Figure 1. Spectral and temporal manifestations of Rydberg blockades. a-c** Differential transmission spectra, $\Delta T/T$, as a function of pump-probe delay ($\Delta t$) measured at variable pump fluences of $f$ = 0.07, 1.01, and 8.15 µJ/cm$^2$, respectively. Throughout, $\Delta t$ is plotted on a logarithmic scale. The pseudocolor scale indicates the $\Delta T/T$ signal, where $T$ is transmission without the pump and $\Delta T$ is pump-induced change in transmission. Red (positive) $\Delta T/T$ features indicate increased transmission (photobleaching), while blue (negative) $\Delta T/T$ features indicate a decreased transmission (photoinduced absorption). **d** Fluence-dependent spectral line cuts at early time delay ($\Delta t$ = 1 ps), where the progression from light green to dark blue traces corresponds to increasing pump fluences: $f$ = 0.07, 0.27, 0.60, 1.51, 3.06, 6.11, and 8.15 µJ/cm$^2$. **e** Time-dependent spectral cuts at a fixed fluence (3.06 µJ/cm$^2$), where the progression of dark purple to light orange corresponds to increased pump-probe delays: $\Delta t$ = 5, 10, 31, 103, 525, 1057, and 1500 ps. All experiments were performed at 7 K. **f** Schematic illustration of the spectral signature of a Rydberg blockade, where the blue-shifted probe resonance, induced by repulsive exciton-exciton interactions ($V(r)$), results in a derivative-like $\Delta T/T$ line shape. A cartoon of the resonant blockade mechanism (i.e., dipolar interaction between excitons of the same $n$) is shown on the right.

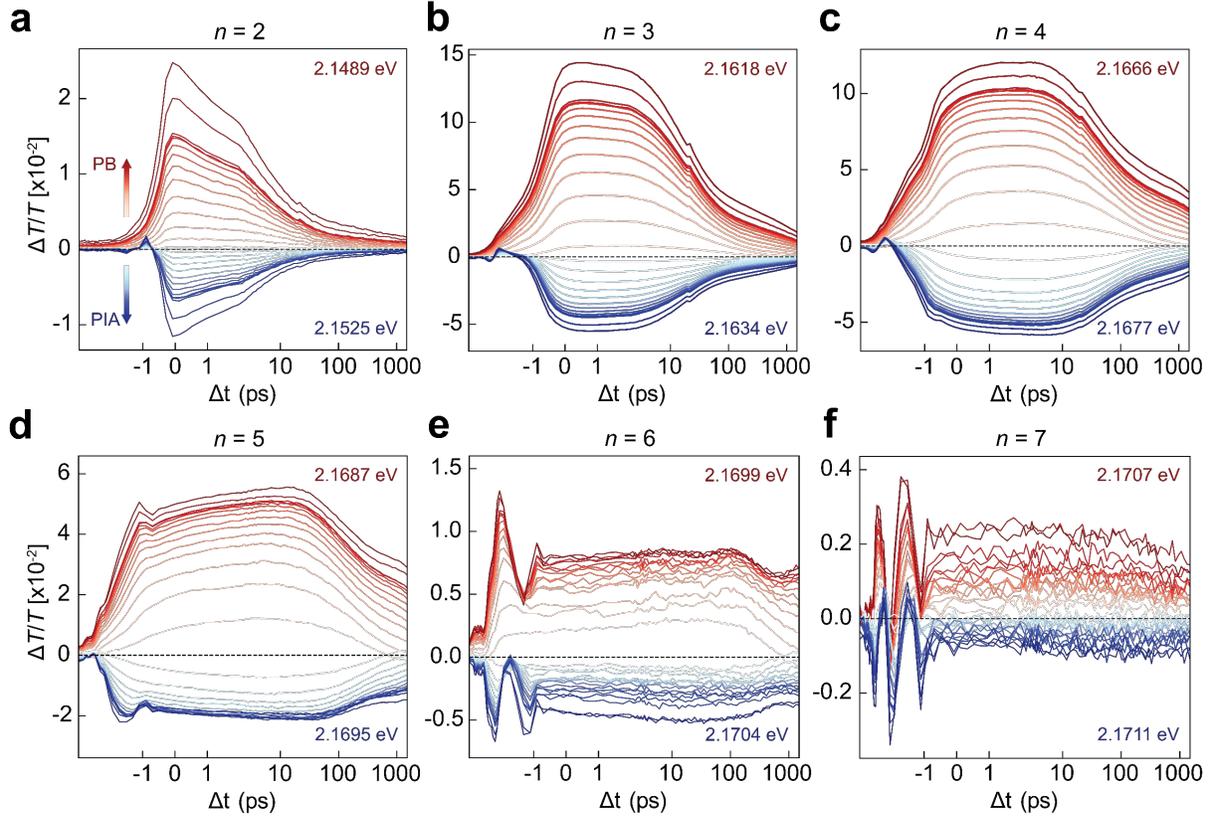

**Figure 2. Fluence-dependent dynamics for *n* = 2-7 as a function of waiting time. a-f** Time traces at fixed probe energies corresponding to the maximum amplitude of the photobleach (PB; red, positive) and photoinduced absorption (PIA; blue, negative) features for each Rydberg state from *n* = 2 to 7. Probe energies used for the PB and PIA traces are indicated in the right-hand side of each panel (PB in the upper corner, PIA in the lower corner). In each panel, the fluence-dependent PB (PIA) time traces are shown from light red (blue) to dark red (blue), with darker color corresponding to increasing pump fluence. Throughout, Δt is plotted on a logarithmic scale. All experiments were performed at 7 K. See Appendix A for further details.

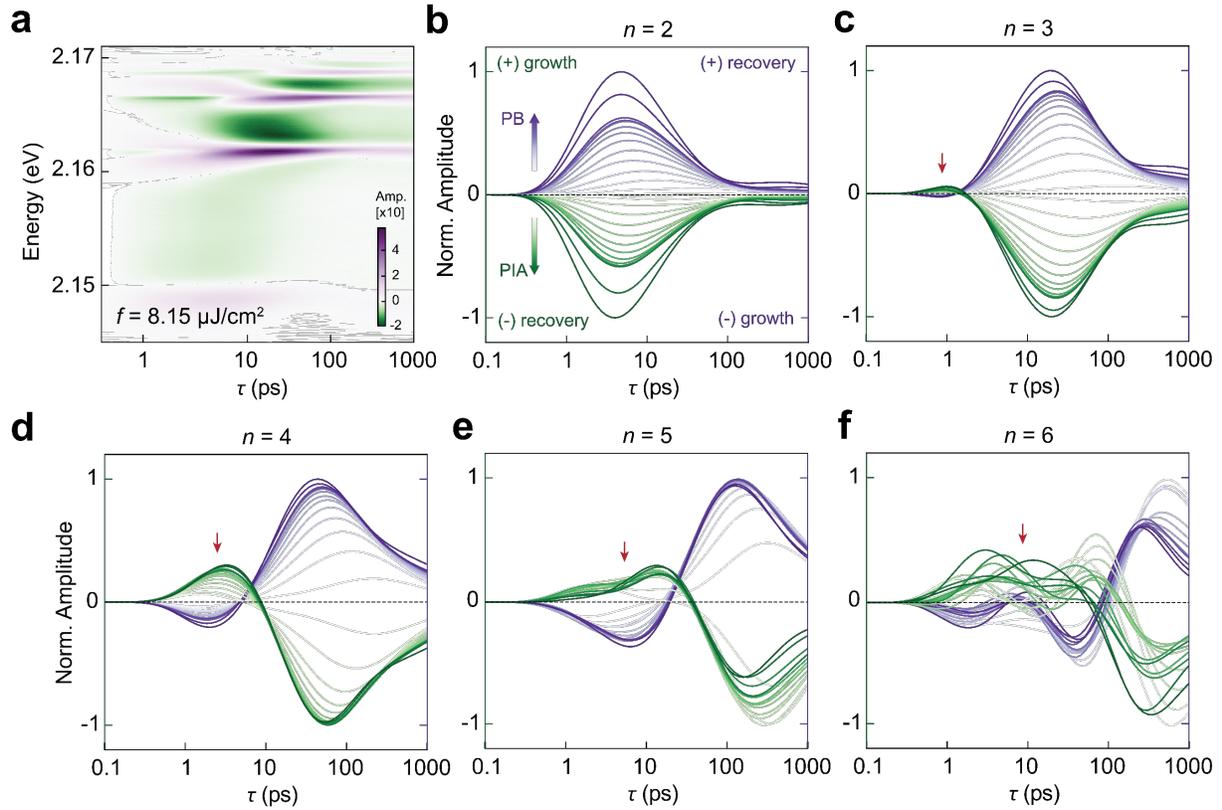

**Figure 3. Fluence-dependent dynamics for *n* = 2-7 as a function of lifetime extracted via LDA.**
**a** Lifetime distribution map for the *f* = 8.15 µJ/cm² data. **b-f** Normalized slices along the fitted τ axis at the probe energies corresponding to the PB and PIA features for each Rydberg level, as described in Fig. 2, for *n* = 2-6, respectively. Here, purple (green) indicated a PB (PIA), where the color change is indicative of these plots being a function of lifetime rather than pump-probe waiting time. For the PB (purple), a positive amplitude indicates signal recovery, while a negative amplitude indicates growth. Conversely, for the PIA (green), a positive amplitude corresponds to growth, and a negative amplitude corresponds to recovery. These conventions are labeled accordingly in the corner of each panel. Red arrows highlight the presence of growths in the PB signals as discussed in the text. See Appendix A for further details.

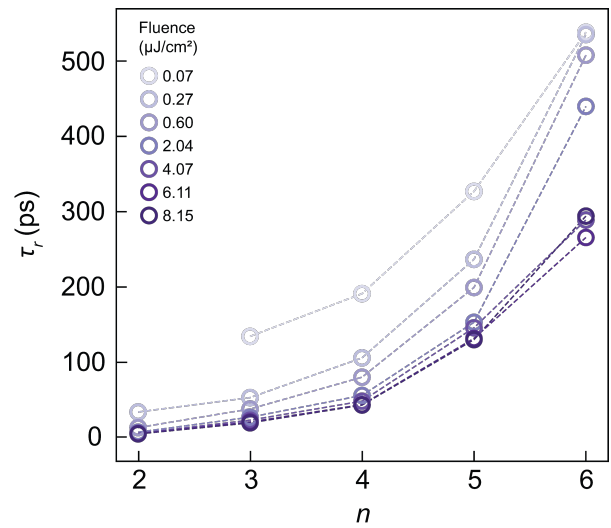

**Figure 4. Fluence-dependent Rydberg exciton recombination times in the blockade regime.** Rydberg exciton recombination time constants, $\tau_r$, versus principal quantum number $n$. Throughout, this trend is shown as a function of fluence where darker shading corresponds to increasing pump fluence, as labeled in the left-hand corner of the plot. The exciton recombination time constants shown here were extracted from the corresponding PB peaks in the LDA results. See Appendix A for further details.

**Appendix A: Methods and Data Analysis**

**Sample Preparation**

The samples are prepared from natural $Cu_2O$ crystals from the Tsumeb mine in Namibia. Pieces of single crystals are removed from the solid matrix and attached with wax to a mounting stub perpendicular to the [111] crystallographic plane. Each crystal is manually polished on both sides, first with 120-grit sandpaper then with 0.5-μm alumina on a polishing cloth. The crystal samples are 3 x 3mm$^2$ in lateral dimension and ~30 μm in thickness. Each sample is placed over a pinhole (1 mm) of a tungsten substrate with copper spacers on quartz windows for transmission experiments. The sample is mounted strain-free by a small quantity of silver paint.

**Spectroscopic Measurements**

For the spectroscopic measurements, the sample is cooled to $T$ = 7 K under vacuum (<10$^{-6}$ torr) in a closed-cycle liquid helium cryostat (Fusion X-Plane, Montana Instruments). Pump-probe measurements are performed using femtosecond pulses (400 kHz, 1030 nm, 250 fs) generated by a solid-state laser (Carbide, Light Conversion). The laser output is split into two beams to form the pump and probe arms. For the pump, we generate the second harmonic at $h\nu_1$ = 2.33 eV, which lies above the optical bandgap of $Cu_2O$ ($E_g$ = 2.17 eV). The pump beam is routed through a motorized delay stage to control the pump-probe time delay (Δt) and modulated using an optical chopper operating at 23 Hz to generate alternating pump-on and pump-off signals. For the probe, the remaining fundamental is focused into a yttrium aluminum garnet (YAG) crystal to generate a stable white-light continuum, which is subsequently filtered to 550-600 nm ($h\nu_2$ = 2.11-2.18 eV) to cover the Rydberg exciton $p$-series of $Cu_2O$. The pump and probe beams are directed non-collinearly onto the sample using an off-axis parabolic mirror, with spot diameters of approximately 250 μm and 100 μm, respectively. The beams are then collected and recollimated using a second off-axis parabolic mirror. The pump arm is spatially filtered, and the probe is dispersed onto an InGaAs detector array (PyLoN-IR, Princeton Instruments). A 550 nm short pass filter is placed before the spectrometer input to block scattered pump light. Pump-on and pump-off spectra at varying pump-probe delays (Δt) are used to calculate the transient transmission signal ($\Delta T/T_{off}$) where $\Delta T = T_{on} - T_{off}$ and the subscript refer to either pump-on or -off.

**Data Analysis**

In this section, we describe the data processing used to analyze the overall transient signal and determine the fluence-dependent exciton (or blockade) lifetimes.

**Determination of Peak Positions.** To identify the peak positions along which to take transient cuts, we interpolate the maximum absolute value of the PB and PIA features for each Rydberg level. In Fig. 2 we take time slices at these energies for each fluence dataset, averaging 2-5 neighboring pixels to capture the entire peak and improve signal-to-noise.

**Lifetime Density Analysis.** Here, we discuss our processing of the transient data with Lifetime Density Analysis (LDA) in detail, including a brief explanation of the approach, as well as a description of the selection of key input parameters.

The underlying principle of LDA is to fit a temporally-resolved dataset to ~100s of exponential decays, returning a semi-continuous distribution of time constants (i.e., lifetimes). The LDA converts the pump-probe time delay Δt, shown below as $t$ for simplicity, into a distribution of lifetimes by fitting the raw transient cuts at each energy $\Delta A(t, \lambda)$ to a convolution of the instrument response function (*IRF*) and $n$ weighted exponentials [30]:

$$\Delta A(t, \lambda) = \sum_{j=1}^{n} x_j(\tau_j, \lambda) e^{-\frac{t}{\tau_j}} \otimes IRF(t, \lambda) \qquad (1)$$

where each exponential $e^{-\frac{t}{\tau_j}}$ corresponds to lifetime $\tau_j$ weighted by an amplitude $x_j(\tau_j, \lambda)$. This amplitude represents the relative contribution of each time constant $\tau_j$ to the dynamics at a given spectral frequency $\lambda$. Peaks in the distribution therefore represent a dominant growth or recombination channel. In our analysis, we fit to $n$ = 200 exponentials distributed logarithmically across the $t$ axis to reflect the raw data sampling. To avoid over-fitting the data, a problem that often arises when dealing with lifetime distributions, we implement Tikhonov regularization with an $\alpha$-value of 3.0, known as the regularization hyper-parameter [35–38]. Here, $\alpha$ is a flexible input parameter that balances the tradeoff between enhancing the distribution coefficients $x_j(\tau_j, \lambda)$ and over-fitting the residuals. The PyLDM software provides statistical measures such as Generalized Cross Validation (GCV), $C_p$, and L-curve statistics to identify the optimal range for $\alpha$ based on the data. We conservatively select $\alpha$ = 3.0, slightly above the predicted safe range ($\alpha$ =

0.01~0.1) to avoid the aforementioned artifacts. To further verify this choice, we repeated the same analysis (LDA, normal fitting, and extraction of trends for fluence-dependent data) with $\alpha$-values 0.3, 1.0, and 5.0 and recovered the same trends as shown in Figs. 4 and B3. Further, input $\tau$ ranges were tested for convergence to ensure no data-coverage issues. The output of the LDA for each fluence is shown in Fig. B2, where the new color scheme distinguishes the LDA results from the raw data. In these pseudocolor plots, purple and green colors correspond to positive and negative amplitudes $x_j(\tau_j, \lambda)$ respectively.

From here, our analysis focuses on the shared *f*- and *n*- dependence of exciton and blockade dynamics. To do this, we take the peak center to be the dominant time constant $\tau_r$ for each LDM cut (Fig. 3). To carefully extract $\tau_r$, we take the logarithm of the $\tau$-axis then fit the normalized amplitude versus $\log(\tau)$ with the normal distribution $A \exp[-((x - x_0)/\sigma)^2/2]$ where $A$ is the amplitude, $\sigma$ is the width parameter, $x$ is $\log(\tau)$, and $x_0$ gives the peak center, $\log(\tau_r)$, from which $\tau_r$ is easily recovered by exponentiation. We use a height threshold to fit top 50% of the peak with the baseline defined by the plateau in the long-time limit. We then plot $\tau_r$ as a function of *n* for the exciton recombination (Fig. 4) and the blockade lifetime (Fig. B3), where each shade corresponds to a different fluence. Error bars are given by the standard deviation of $\sigma$ in the 99% confidence interval, however, since LDM cuts are pre-smoothed by the regularization procedure, resulting error in the fits is negligibly small. We thus omit error bars for the extracted $\tau_r$ in Fig. 4 and Fig. B3.

**Appendix B: Extended Data**

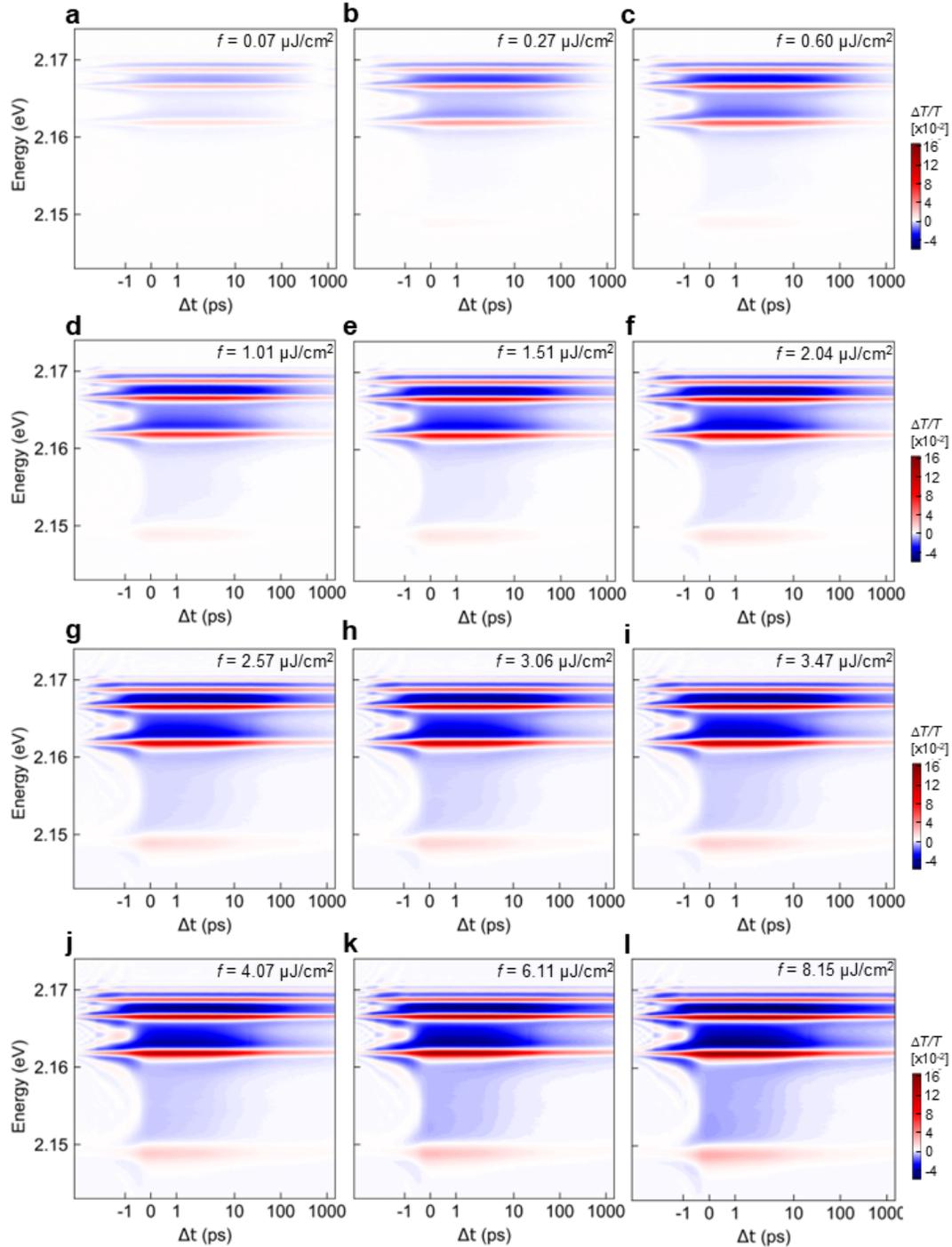

**Figure B1. Transient spectra at all measured fluences. a-l** Differential transmission spectra, *ΔT/T*, as a function of pump-probe delay (Δt) at all measured fluences from *f* = 0.07 to 8.15 µJ/cm². Throughout, Δt is plotted on a logarithmic scale. As in the main text, the pseudocolor scale indicates the *ΔT/T* signal, where *T* is transmission without the pump and *ΔT* is pump-induced change in transmission. Red (positive) *ΔT/T* features indicate increased transmission (photobleaching), while blue (negative) *ΔT/T* features indicate a decreased transmission (photoinduced absorption). All data collected at T = 7 K.

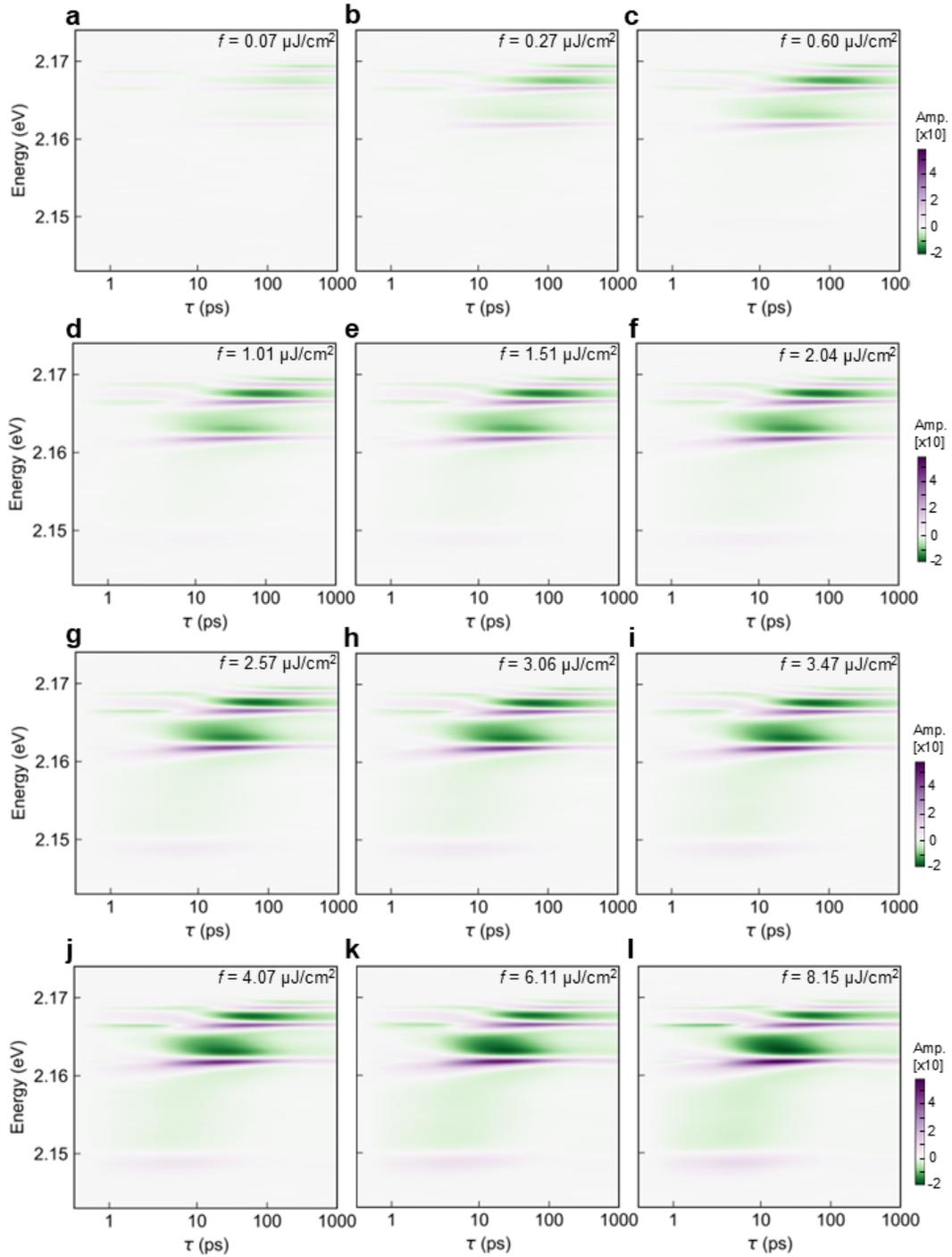

**Figure B2. Lifetime Density Maps (LDMs) for all measured fluences. a-l** Output of Lifetime Density Analysis (LDA) performed for all fluence-dependent datasets, with the $\tau$-axis plotted in logscale. An $\alpha$-value of 3 was used to narrow the distribution without over-fitting the residuals. The new color scheme distinguishes the LDA results from the raw data in Fig. B1. In these pseudocolor plots, purple and green colors correspond to positive and negative probability amplitudes respectively.

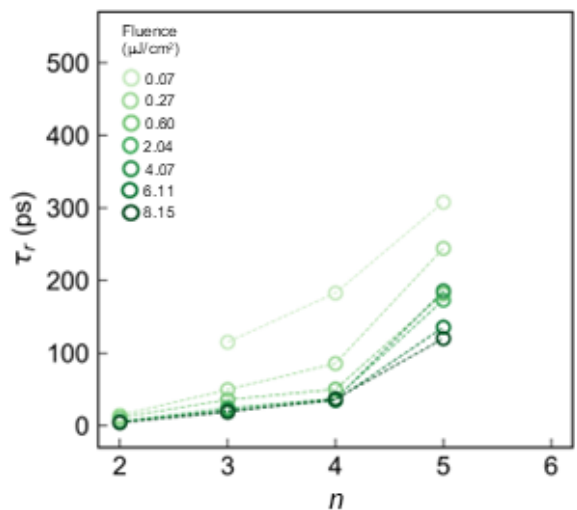

**Figure B3. Fluence-dependent Rydberg blockade lifetimes.** Rydberg blockade decay time constants, $\tau_r$, versus principal quantum number $n$. Throughout, this trend is shown as a function of fluence where darker shading corresponds to increasing pump fluence, as labeled in the left-hand corner of the plot. The blockade recovery time constants shown here were extracted from the corresponding PIA peaks in the LDA results. We do not analyze LDA results for $n = 6$ for PIA due to insufficient signal-to-noise.